\documentclass[reprint,superscriptaddress,amsmath,amssymb,aps,prb,
]{revtex4-2}

\usepackage{graphicx}% Include figure files
\usepackage{dcolumn}% Align table columns on decimal point
\usepackage{bm}% bold math
\usepackage[colorlinks=true, linkcolor=blue, citecolor=blue, urlcolor=blue, hypertexnames=false]{hyperref}

\begin{document}
\title{Symmetry-breaking induced surface magnetization in non-magnetic RuO$_2$}
% \author{Me}
% \email{mail@example.com}
% \author{Myself}
% \author{Someone Else} you 
% \affiliation{A University}

\author{Dai Q. Ho}
\email{daiqho@udel.edu}
\affiliation{Department of Materials Science and Engineering, University of Delaware, Newark, Delaware 19716, United States}
\affiliation{Faculty of Natural Sciences, Quy Nhon University, Quy Nhon 55113, Vietnam}

\author{D. Quang To}
\affiliation{Department of Materials Science and Engineering, University of Delaware, Newark, Delaware 19716, United States}

\author{Ruiqi Hu}
\affiliation{Department of Materials Science and Engineering, University of Delaware, Newark, Delaware 19716, United States}

\author{Garnett W. Bryant}
\email{garnett.bryant@nist.gov}
\affiliation{Nanoscale Device Characterization Division, Joint Quantum Institute, National Institute of Standards and Technology, Gaithersburg, Maryland 20899-8423, United States}
\affiliation{University of Maryland, College Park, Maryland 20742, USA}

\author{Anderson Janotti}
\email{janotti@udel.edu}
\affiliation{Department of Materials Science and Engineering, University of Delaware, Newark, Delaware 19716, United States}

\date{\today}% It is always \today, today,
  
\begin{abstract}
Altermagnetism is a newly identified phase of magnetism distinct from ferromagnetism and antiferromagnetism. RuO\texorpdfstring{$_2$}{2} has been considered a prototypical metallic altermagnet with a critical temperature higher than room temperature. Previous interpretations of unusual magnetic properties of RuO\texorpdfstring{$_2$}{2} relied on the theoretical prediction that local moments on two Ru sublattices, which are connected by four-fold rotational symmetry, are quite significant (\texorpdfstring{$\sim1\,\mu_B$}{~1 muB}), leading to long-range antiferromagnetic order. However, accumulated experimental data suggest that local moments on Ru in RuO\texorpdfstring{$_2$}{2} are vanishingly small, indicating the bulk material is likely non-magnetic. This observation is consistent with the delocalized nature of the \texorpdfstring{$4d$}{4d} electron of Ru and the strong screening effect in the metallic state. In this work, we show that despite the non-magnetic ground state in the bulk, RuO\texorpdfstring{$_2$}{2}(110) surface exhibits spontaneous magnetization, which we attribute to the breaking of local symmetry, resulting in electronic redistribution and magnetic moment enhancement. The emergence of surface magnetism gives rise to interesting spectroscopic phenomena, including spin-polarized surface states, spin-polarized scanning probe microscopy images, and potentially spin-dependent transport effects. These highlight the important role of surface magnetic structures in the otherwise non-magnetic bulk RuO\texorpdfstring{$_2$}{2}.
\end{abstract}

\maketitle

%%%%%%%%%%%%%%%%%%%%%%%%%%%%%%%%%%%%%%%%%%%%%%%%%%%%%%%%%%%%%%%%%%%%%%%%%%%%%%%%%%%%%%%%%%%%%%%%%%%%%%%%%%%%%%%%%%%%%%%%%%%%%%
\section{Introduction} \label{sec:intro}

Altermagnetism (AM) is an emerging phase of magnetism that differs from the well-known ferromagnetism (FM) and antiferromagnetism (AFM) ~\cite{ref1vsmejkal2022beyond}. In altermagnetic materials, spin-splitting in the electronic band structure occurs alternately in reciprocal space directions that are related to each other by rotation or reflection but not translation or inversion, which is determined by the presence of non-magnetic ligands in real space~\cite{ref1vsmejkal2022beyond}. Among the earliest and most studied candidates for altermagnetic materials is RuO$_2$, which has been considered a prototype altermagnet~\cite{ref1vsmejkal2022beyond,ref2vsmejkal2020crystal}. Despite various theoretically predicted~\cite{ref2vsmejkal2020crystal} and experimentally observed phenomena linked to time-reversal symmetry breaking in AM such as the anomalous Hall effect~\cite{ref3feng2022anomalous}, spin-splitter effect~\cite{karube2022observation}, magnetic circular dichroism~\cite{ref4fedchenko2024observation}, ascribed to the assumption that RuO$_2$ is an AM with a high N\'eel temperature and a significant local moment (typically assumed $\sim1\,\mu_B$ in theoretical calculations), direct experimental evidence for such a high moment in this $4d$ metallic oxide is lacking. Therefore, the magnetic configuration of RuO$_2$ and especially the presence of local moments and long-range magnetic ordering are still subjects of debate \cite{smolyanyuk2024fragility}. The first neutron scattering experiment on RuO$_2$ reported a local moment of only \(0.05\,\mu_B\)~\cite{ref5berlijn2017itinerant}. Subsequently, an experiment using the resonant X-ray scattering technique supported the AFM ground state in RuO$_2$~\cite{ref6zhu2019anomalous}. However, recent in-depth investigations strongly suggest that the ground state of RuO$_2$ is non-magnetic/paramagnetic~\cite{ref7hiraishi2024nonmagnetic,ref8liu2024,ref9Kessler2024,kiefer2025crystal,wenzel2025fermi,osumi2025spin}. For instance, local-moment-sensitive muon spin rotation ($\mu$SR) and neutron diffraction experiments have shown that the ground state of the bulk RuO$_2$ is likely non-magnetic, with negligibly small local moments of \(10^{-4} \,\mu_B\)~\cite{ref7hiraishi2024nonmagnetic,ref9Kessler2024}. Furthermore, by using broadband infrared spectroscopy combined with first-principles calculations to probe the optical conductivity of RuO$_2$---a bulk electronic property---researchers concluded that bulk RuO$_2$ is well described by a non-magnetic model~\cite{wenzel2025fermi}. Very recently, Liu {\rm et al.}\ have directly probed the electronic structure of RuO$_2$ in bulk and thin film samples using angle-resolved photoemission spectroscopy (S-ARPES) and found no evidence of spin splitting in the electronic structure of the material~\cite{ref8liu2024}. These findings align with the delocalization and weak correlation nature of the Ru $4d$ orbitals, consistent with the metallic property of the compound.

Most of the experiments mentioned above have been concentrated on bulk or thick films, with less focus on surface properties. However, the surface properties of altermagnets might play a significant role in various phenomena at surfaces and interfaces or even be key to explaining some observations~\cite{ref12Sattigeri2023,ref13Chilcote2024}. Of many low-index surfaces, RuO$_2$ surface and thin films in the crystallographic direction $\langle 110 \rangle$ are of particular interest since that is the crystal surface with the lowest surface energy and is of easy cleavage~\cite{ref14Over2012,ref15Xu2014}. For example, the RuO$_2$(110) surface has been a primary focus in catalysis research thanks to its high electrocatalytic water splitting activity, attributed to the presence of Ru dangling bonds~\cite{ref14Over2012,ref16Over2000}. However, most studies have ignored the magnetic properties of the surface, except for a couple of theoretical predictions, which highlighted the role of surface magnetism in RuO$_2$(110) in catalytic reactions during electrolysis~\cite{ref17Torun2013,liang2022anti}. Given the subtlety of magnetism in RuO$_2$ seen in the literature, it is possible that surface effect could be a relevant source of magnetic phenomena observed in various samples. For instance, strained RuO$_2$(110) thin films grown on TiO\(_2\) substrates have recently been shown to exhibit superconducting behavior at low temperatures and a metal-to-insulator transition in the ultrathin limit regime~\cite{ref18Uchida2020,ref19Ruf2021,ref20Rajapitamahuni2024}. The role of the soft-phonon mode and the enhancement of electron density at the Fermi level ($E_{\text{F}}$) have been invoked to explain this superconductivity. However, the possible role of the magnetic structure in supporting these phenomena has been overlooked.

As noted previously, Torun {\rm et al.}~\cite{ref17Torun2013} have predicted that surface magnetization can spontaneously develop on the RuO$_2$(110) surface. However, their study was not without limitations. First, the slab used in their calculations, consisting of only five layers of RuO$_2$, was probably insufficient to accurately represent both surface and bulk-like regions, potentially affecting the predicted behavior of surface magnetization. In metallic systems such as RuO$_2$, the electronic wavefunctions of the top and bottom surfaces of the slab used in the calculation can penetrate deep into the bulk and mutually interact if the slab is not sufficiently thick, leading to an alteration of electronic properties of the true surface. Second, the origin of the surface magnetism was not comprehensively explored. More importantly, since their work focused on surface catalysis, it did not address how surface magnetization influences electronic and magnetic properties that are critical for spectroscopic and transport phenomena, which is highly relevant for spintronics applications~\cite{ref4fedchenko2024observation,ref21Jiang2023}.

In this paper, using first-principles calculations based on density functional theory (DFT), we show that surface magnetism spontaneously develops in RuO$_2$(110). The spontaneous surface magnetization can be understood from the breaking of symmetry on the surface of RuO$_2$(110), leading to a significant reconstruction of the electronic structure of RuO\(_6\) octahedra at the surface layer, i.e., in the band filling of the Ru $4d$ orbitals, resulting in sizable local moments. Remarkably, the presence of spontaneous surface magnetization leads to spin-polarized surface states and spin-dependent transport effects. These results can be important for understanding the experimental data on RuO$_2$ obtained using spectroscopic and transport measurements such as spin- and angle-resolved photoemission spectroscopy (S-ARPES)~\cite{ref8liu2024,ref22Lin2024}, spin-polarized scanning tunneling microscopy/spectroscopy (SP-STM/SP-STS), anomalous Hall measurements~\cite{ref3feng2022anomalous}, and interfacial spin-dependent transport phenomena~\cite{kobayashi2024detection}.

%%%%%%%%%%%%%%%%%%%%%%%%%%%%%%%%%%%%%%%%%%%%%%%%%%%%%%%%%%%%%%%%%%%%%%%%%%%%%%%%%%%%%%%%%%%%%%%%%%%%%%%%%%%%%%%%%%%%%%%%%%%%%%
\section{Computational Method} \label{sec:Com}

Our calculations are based on DFT~\cite{ref23Hohenberg1964,ref24KohnSham1965} and the projector augmented wave (PAW) method was employed to treat interactions between the valence electrons and the ionic cores~\cite{ref27_Blochl1994PAW,ref28kresse1999ultrasoft} as implemented in the \textsc{VASP} code~\cite{ref25vasp1996CMS,ref26vasp1996PRB, NISTdisclaimer}.  We used the recommended standard PAW potentials for Ru and O, \texttt{Ru\_pv} (4\(p^6\)4\(d^7\)5\(s^1\)) and \texttt{O} (2\(s^2\)2\(p^4\)). Bloch wavefunctions of the materials were expanded in a plane-wave basis set with a cutoff energy of 600 eV. Due to the metallic nature of RuO$_2$, the Methfessel-Paxton smearing method (\texttt{ISMEAR} = 1) in combination with a smearing value of 0.2 eV was used for integration over the Brillouin zone (BZ) when optimizing structural parameters. For very accurate total energy calculations, we utilized the tetrahedron method with Bl\"ochl corrections (\texttt{ISMEAR} = \(-\)5) and a Gamma-centered k-point mesh to sample the BZ. Data postprocessing made use of \textsc{Vaspkit}~\cite{WANG2021108033}.

Bulk RuO$_2$ was simulated using a tetragonal unit cell with a rutile structure (\(P4_2/mnm\), SG 136). Due to the metallic property of bulk RuO$_2$, a dense $k$-point mesh of 12×12×18 was used to sample the bulk BZ. The lattice constant was adopted from a recent experimental determination~\cite{ref29Burnett2020}. RuO$_2$ films along [110] were built by reorienting the lattice vectors of the primitive cell, i.e., primitive lattice vectors along [001], [\(1\overline{1}0\)], and [110] become [100], [010], and [001] of the RuO$_2$(110) surface unit cell, respectively, thus giving us a nonpolar stoichiometric slab. The symmetry of the bulk unit cell transitions to \(Pmmm\) (SG 47) for stoichiometric slabs with an odd number of Ru layers. A $k$-point mesh of 12×6×1 was used for structural optimization of the surface unit cell, and a denser mesh of 18×9×1 was used for density of states (DOS) calculations. To eliminate spurious image interactions between slabs, a vacuum space of at least 20 \AA~was added to the slab normal direction. Internal coordinates of the slab were fully relaxed until the Hellman-Feynman force on each atom was smaller than 0.005 eV/\AA~ while keeping the lattice vectors fixed to the experimental bulk values. Due to the recent experimental observation of a non-magnetic ground state in bulk RuO$_2$~\cite{ref7hiraishi2024nonmagnetic,ref8liu2024,ref9Kessler2024,kiefer2025crystal,wenzel2025fermi,osumi2025spin}, we employed the PBE+$U$ approach with $U = 0$~\cite{ref30Perdew1996,ref31Perdew1997,dudarev1998electron} to describe its electronic structure. Additionally, since spin-orbit coupling does not significantly affect the electronic and magnetic structures of RuO$_2$, it was neglected in our calculations.

To gain insights into the chemical bonding between atomic pairs contributing to the emergence of spin-polarized states at the surface of RuO$_2$, we performed a crystal orbital Hamilton population (COHP) analysis~\cite{ref32Dronskowski1993,ref33Deringer2011}. The COHP method decomposes the one-particle band energies into interactions between atomic orbitals of adjacent atoms. It effectively weighs the DOS by the corresponding Hamiltonian matrix elements, thus recovering the phase information of calculated wavefunctions that is otherwise missing in the band structure or DOS descriptions of material electronic structures. This enables the identification of bonding, non-bonding, and antibonding characteristics among the pairwise atoms. The stability of these interactions is quantified by the COHP values. A positive COHP value indicates bonding interactions, while a negative value represents an antibonding character. Traditionally, COHP analysis was done directly from atom-centered basis set calculations. However, since our calculations were done using the plane-wave basis set, a projection scheme from this basis set to a local orbital basis set was employed using the \texttt{LOBSTER} code~\cite{ref32Dronskowski1993,ref33Deringer2011,ref34Nelson2020}. In this projection scheme, the equivalence of the traditional COHP quantity is \(-\)pCOHP (negative of projected COHP). Our study applied COHP analysis to examine the bonding character and stability of the nearest-neighboring atomic pairs around Ru atoms on the surface and in bulk-like regions. For the projection, we employed the local basis function as defined in \texttt{pbeVaspFit2015}.

Since spin-polarized scanning tunneling microscopy (SP-STM) can be used to recognize spin polarization effects on the surface~\cite{ref35Wortmann2001}, we simulated spin-resolved STM images employing the Tersoff-Hamann approximation~\cite{ref36Tersoff1985}. In this approach, the tunneling current at the simulated probe tip (i.e., at a particular distance from the surface) in an STM experiment is proportional to the local density of states (LDOS) of the integrated electronic states ranging from the Fermi level to a predefined energy level given by a bias voltage \(V\), roughly corresponding to an experimental voltage. The LDOS is given by: 
\begin{equation}
n(r, E) = \sum_{i} \left| \psi_{i}(r) \right|^2 \delta \left( \varepsilon_{i} - E \right),  
\label{eqn1}
\end{equation}  
and the tunneling current can be expressed as  

\begin{equation}
I(r, V) \propto \int_{E_{\mathrm{F}}}^{E_{\mathrm{F}} + eV} n(r, E) \, dE,  
\label{eqn2}
\end{equation}
where \(n(r, E)\) represents the LDOS at a given position \(r\) and energy \(E\). The LDOS can be evaluated from partial charge densities calculated from a pre-converged wavefunction. The partial charge density file from VASP was read by the \texttt{HIVE-STM} program~\cite{ref37Vanpoucke2008}, and STM images were generated using the constant-height method (at a height of \(\sim 3\)~\AA{} above the highest atoms of the slab---the oxygen bridging atoms). For a given bias \(V\), the simulated STM images reflect the contrast in the partial charge densities within the energy range $0<E-E_{\text{F}}<V$ (for \(V > 0\), positive bias) and $V<E-E_{\text{F}}<0$ (for \(V < 0\) for negative bias).

%%%%%%%%%%%%%%%%%%%%%%%%%%%%%%%%%%%%%%%%%%%%%%%%%%%%%%%%%%%%%%%%%%%%%%%%%%%%%%%%%%%%%%%%%%%%%%%%%%%%%%%%%%%%%%%%%%%%%%%%%%%%%%
\section{Results and Discussion}\label{sec:result}
%%%%%%%%%%%%%%%%%%%%%%%%%%%%%%%%%%%%%%%%%%%%%%%%%%%%%%%%%%%%%%%%%%%%%%%%%%%%%%%%%%%%%%%%%%%%%%%%%%%%%%%%%%%%%%%%%%%%%%%%%%%%%%
%\subsection{Electronic and magnetic properties of bulk RuO$_2$}
\subsection{Electronic and magnetic properties of bulk RuO\texorpdfstring{$_2$}{2}}

\begin{figure*}
\centering
\includegraphics[width=7.0 in]{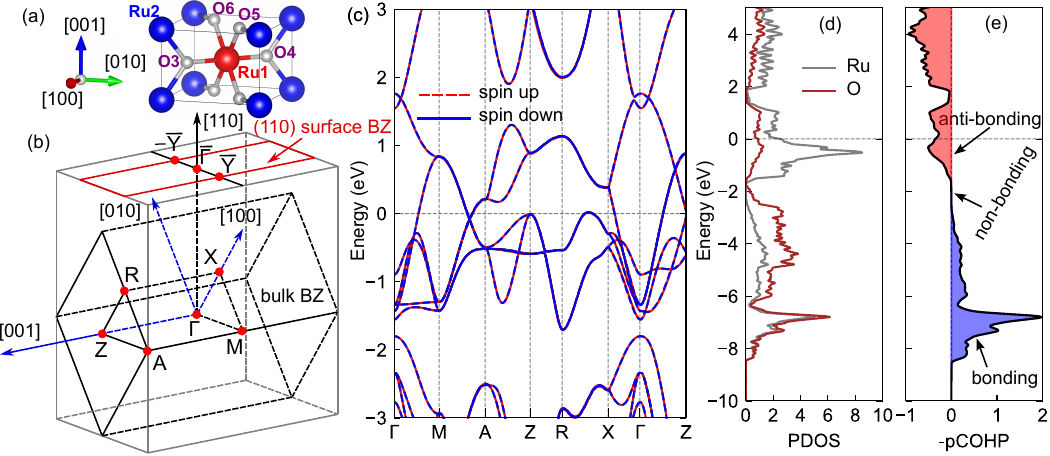}
\caption{Rutile-type crystal structure, electronic structure and chemical bonding analysis of bulk RuO$_2$. (a) Rutile structure of RuO$_2$ with two Ru sublattices in red and blue, oxygen in silver color, (b) Bulk Brillouin zone (BZ) shown in black and rotated so that [110] crystallographic direction normal to the (110) surface whose BZ shown in red, (c) spin-polarized electronic band structure of the bulk exhibiting no spin-polarization ($E_{\text{F}}$ set to 0), (d) projected density of states on Ru and O in the bulk region, and (e) the corresponding \(-\)pCOHP curve on the same energy scale.}
\label{fig1}
\end{figure*}

RuO$_2$ crystallizes in the tetragonal rutile structure. Its bulk unit cell consists of two formula units, as shown in Fig.~\ref{fig1}(a). The two Ru atoms reside at the center (\( \frac{1}{2}, \frac{1}{2}, \frac{1}{2} \)) and corner (0, 0, 0) of the unit cell, and are coordinated octahedrally distorted by the surrounding O atoms, forming two Ru sublattices. RuO$_2$ belongs to the centrosymmetric tetragonal structure described by space group \(P4_2/mnm\) with the point group \(4/mmm\). Each Ru resides at the center of a distorted octahedron, with the octahedra being interconvertible through a \(C_4\) rotational axis aligned along the [001] crystallographic direction.

The spin-resolved electronic band structure of the bulk is shown in Fig.~\ref{fig1}(c), which does not exhibit spin polarization. In the assumed altermagnetic state, the electronic structure of interest lying along the $\Gamma$--$\mathrm{M}$ high-symmetry path of the bulk Brillouin zone (BZ) shows significant spin splitting (\(> 1\) eV) as described by theoretical calculations of the band structure with relatively large on-site Hubbard \(U\) values (\(U \sim 1.5 - 2\) eV)~\cite{ref2vsmejkal2020crystal,ref5berlijn2017itinerant}. However, this spin splitting does not appear in the non-magnetic state. Along this momentum direction, three bands are present near the Fermi level, with an additional flat band lying around \(-1.5\) eV below the Fermi level. When considering the electronic structure of the (110) slab, the $\Gamma$--$\mathrm{M}$ path of the bulk BZ is of particular importance since bulk band dispersions along this path and parallel to it are projected onto the high-symmetry path $\overline{\Gamma}$--$\overline{\mathrm{Y}}$ of the surface BZ, as illustrated in Fig.~\ref{fig1}(b).

The DOS and chemical bonding analysis, shown in Fig.~\ref{fig1}(d) and ~\ref{fig1}(e), highlights important characteristics of Ru--O bonds in RuO$_2$. The DOS in Fig.~\ref{fig1}(d) clearly shows strong hybridization between the Ru and O orbitals over a wide range of energy. Notably, between 8 and 6 eV below the Fermi level, pronounced peaks in the atomic-projected density of states (PDOS) and \(-\)pCOHP (positive---bonding character) with similar contributions from both Ru and O indicate a strong covalent bonding, best described by the \(\sigma\)-bonding composed of Ru $4d$ (\(e_g\)) and O $2p$ orbitals~\cite{ref38Sorantin1992}. In addition, approximately from 6 eV to 3 eV below the Fermi level, small positive values of \(-\)pCOHP, with a greater contribution of oxygen, suggest a weaker covalent bonding character of the Ru--O bonds due to the \(\pi\)-bonding between Ru $4d$ (\(t_{2g}\)) and O $2p$. Around \(-2\) eV, the contribution of Ru disappears, i.e., Ru PDOS is zero, leaving only O $2p$ states. This indicates that the electrons occupying these states belong to non-bonding orbitals and do not contribute to the Ru--O bonding strength, leading the \(-\)pCOHP values to approach zero.

Interestingly, in the vicinity of the Fermi level, i.e., from \(-2\) eV to 2 eV, the contribution of Ru $4d$ becomes dominant and populate antibonding crystal orbitals, as indicated by negative \(-\)pCOHP values. This implies potential instability when electrons occupy these states. Fortunately, the DOS and \(-\)pCOHP peaks lie approximately 0.5 eV below the Fermi level, alleviating the instability due to filling antibonding orbitals~\cite{ref39Landrum2000,ref40Dronskowski2004}. Ultimately, despite the presence of antibonding Ru--O interactions near the Fermi level in the \(-\)pCOHP curves, the integrated values of \(-\)pCOHP for Ru--O bonds (\(-\)IpCOHP values) are positive (Table~\ref{tables1}), indicating a net bonding character for these interactions. Consequently, from a chemical perspective, there is no driving force for electronic reconstruction (i.e., the redistribution of the two spin channels), resulting in a stable non-magnetic ground state rather than an ordered magnetic one.

These characteristics suggest strong mixed covalency and ionicity in Ru--O bonds, which is consistent with the atomic charge states of Ru and O obtained from population analysis based on charge density using the Bader charge method via the atom-in-molecule (AIM) approach~\cite{ref41Bader2003,ref42Tang2009} or wavefunction-based methods such as Mulliken~\cite{ref43Mulliken1955} and L\"owdin~\cite{ref44Lowdin1950} charges obtained within the projection scheme of the \texttt{LOBSTER} code~\cite{ref45Ertural2019}. The charge states of the atoms determined from these analyzes are lower in magnitude than the formal values of \(+4\) for Ru and \(-2\) for O, as shown in Table~\ref{table1}.

\begin{table}
\setlength{\tabcolsep}{12pt}
\caption{Atomic charge states of Ru and O in bulk RuO$_2$}
\centering
\begin{tabular}{l c c c}
\hline\hline
Population & Ru & axial-O & equatorial-O\\  
\hline 
Bader & 1.74 & -0.87 & -0.87 \\
Mulliken & 1.38 & -0.69 & -0.69 \\
L\"owdin & 1.20 & -0.60 & -0.60 \\[0.5ex]
\hline\hline %inserts single line
\end{tabular}
\label{table1} % is used to refer this table in the text
\end{table}

Despite the non-magnetic character of bulk RuO$_2$, intriguing phenomena can arise at the (110) surface, leading to the emergence of surface magnetism. This disparity between the bulk and surface properties can be attributed to the altered electronic structure and bonding characteristics at the surface, which disrupt the symmetry and electronic interactions present in the bulk. The unique environment at the surface can facilitate localized magnetic moments that do not exist in the bulk material. In the next section, we will explore how these localized states, influenced by surface geometry and reduced coordination, contribute to the overall magnetic behavior of the (110) surface. 

%%%%%%%%%%%%%%%%%%%%%%%%%%%%%%%%%%%%%%%%%%%%%%%%%%%%%%%%%%%%%%%%%%%%%%%%%%%%%%%%%%%%%%%%%%%%%%%%%%%%%%%%%%%%%%%%%%%%%%%%%%%%%%
%\subsection{Surface magnetization in RuO$_2$(110) surface}
\subsection{Magnetization on RuO\texorpdfstring{$_2$}{2}(110) surface}

\subsubsection{Emergence of surface magnetization}
%\subsubsection{The emergence of surface magnetization in RuO\texorpdfstring{$_2$}{2}(110) surface}

\begin{figure*}
\centering
\includegraphics[width=7.0 in]{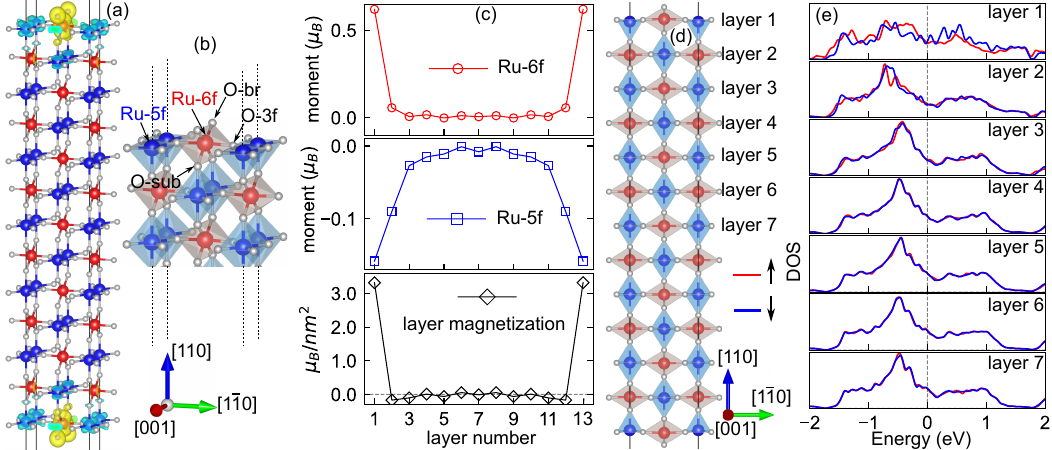}
\caption{Geometrical structure of the (110) surface and the emergence of surface magnetism. (a) The slab structure of 13L RuO$_2$(110) includes two Ru sublattices inherited from the bulk: Ru-center in red, Ru-corner in blue, and oxygen ligands in silver. At the surface layers, Ru-center atoms are fully octahedrally coordinated by six O ligands (denoted as Ru-6f), while the Ru-corner sublattice has one dangling bond due to surface termination (Ru-5f), see (b) for a better view. Also shown is the magnetization density visualization, which exhibits significant accumulation of spin moments only at the outermost layers, with spin-up in yellow and spin-down in cyan. The isosurfaces are plotted using 5\% of the maximum isosurface value. (b) Detailed atomic structure of the surface includes stoichiometric surface termination, with labels for atoms in the surface layer. (c) The spin moment at each Ru site in each layer (top two panels), and the magnetization of each layer (bottom panel) are plotted as functions of distance from the bulk-like region. (d) View along the [001] direction showing the layers numbered for the layer-projected DOS in (e), demonstrating that spin polarization decays quickly when approaching the inner layers.}
\label{fig2}
\end{figure*}
    
To gain insight into the magnetic structure of the RuO$_2$(110) surface, we performed calculations using the slab model with 3D periodic boundary condition. 
%In isostructural insulating materials such as TiO\(_2\), the (110) surface is electrostatically stable. However, electrostatics are not a concern in the case RuO$_2$ since it is a metallic system. 
We simulate a surface with a bulk-like region well represented in the middle of the structure, choosing stoichiometric slabs with an odd number of Ru layers. These slabs possess inversion symmetry, eliminating unnecessary complications related to asymmetric systems, such as spurious charge transfer and build up electric fields across the slab thickness. The slab thickness ranges from three layers (denoted as 3L) up to nineteen layers (19L). Each layer is composed of three atomic planes, with one Ru-containing layer sandwiched between two oxygen-only layers. Each slab has two identical surface layers at the top and bottom of the unit cell [cf. Fig.~\ref{fig2}(a)].

The converged, relaxed thirteen-layer slab unit cell (13L) is shown in Fig.~\ref{fig2}(a). This symmetric structure represents both bulk and surface regions since it is thick enough to decouple the two surfaces, and the middle layer behaves as a bulk-like region [see Fig.~\ref{figs1} in the Supplemental Material (SM) for the dependence of local moments of surface Ru atoms and top-layer magnetization on slab thickness]. Each layer stacked along the [110] direction contains two Ru atoms corresponding to the Ru-center and Ru-corner sublattices of the bulk [cf. Fig.~\ref{fig1}(a)]. This arrangement would yield a magnetically compensated structure with oppositely pointed moments within each (110) plane if the ground state of the bulk was antiferromagnetically ordered with the Néel vector along [001], as assumed earlier---similar to the case of the isostructural FeF\(_2\) antiferromagnet~\cite{ref46Weber2024}. 

Unlike the magnetic structure in traditional antiferromagnets, where magnetization and spin density vanish, the assumed altermagnetic spin structure of RuO$_2$ would yield distinct spin densities associated with opposite spin sublattices. That is, real-space spin densities of sublattices connected by the four-fold rotation symmetry around the [001] axis would differ~\cite{ref1vsmejkal2022beyond}. However, in non-magnetic RuO$_2$ bulk, the absence of local magnetic moments suggests that the antiferromagnetic order within each Ru plane of the (110) slab and the altermagnetic spin structure are unlikely. This is evident in the spin density visualization for the (110) slab displayed in Fig.~\ref{fig2}(a) and the layer dependence of local moments on each Ru site from the surface to the bulk-like region shown in Fig.~\ref{fig2}(c) (top and middle panels). Most atoms in the slab are non-magnetic except for those in the outermost layers.

In contrast to the bulk, local moments develop on the surface layers of the slab, decaying rapidly into the bulk. The collinear moments arrange ferrimagnetically within the surface layers, with unequal spin moments: Ru-center (\(\sim\)0.61 \(\mu_\mathrm{B}\)) and Ru-corner (\(\sim\)\(-0.17\) \(\mu_\mathrm{B}\)) couple collinearly in opposite directions. The majority spin density is located primarily on the Ru-center atom (denoted Ru-6f due to its six-fold coordination) and bridging oxygen (O-br, spin moment \(\sim\)0.19 \(\mu_\mathrm{B}\)), while the Ru-corner atom (denoted Ru-5f due to its five-fold coordination) dominates the minority spin density counterpart [cf. Fig.~\ref{fig2}(b) for labels of surface atoms]. Other oxygen ligands of the outermost layers exhibit non-spin-polarized behavior [Fig.~\ref{fig2}(a)]. Notably, the induced surface magnetization in RuO$_2$(110) reaches a substantial value of $\sim$$3.0\,\mu_{\mathrm{B}}/\mathrm{nm}^2$---two orders of magnitude larger than the predicted magnetization in the isostructural rutile FeF\(_2\)~\cite{ref46Weber2024}, despite the latter exhibiting an intrinsic AFM ordering.

As seen in Fig.~\ref{fig2}(a) and evidenced by variations in the local moments for each Ru sublattice across layers, the magnetization density decreases dramatically from the surface layers to the bulk-like layers. Local moments on Ru, and consequently layer magnetizations, nearly vanish starting from the layer just beneath the surface [Fig.~\ref{fig2}(c), bottom panel]. The emergence of surface-layer magnetism and the rapid decrease in magnetization can also be observed in the layer-projected DOS in Fig.~\ref{fig2}(e), where significant spin splitting is shown to occur mainly on the top layer. This magnetic structure is lower in energy than the non-magnetic counterpart by $\sim$45 meV per unit cell, independent of the slab thickness, consistent with the origin of the magnetic structure being solely due to the surface layer (see Fig.~\ref{figs2} for the relative energy of the spin-polarized state compared to the non-spin-polarized counterpart as a function of slab thickness).

In a recent study, Weber {\em et al.}~\cite{ref46Weber2024} demonstrated that surface magnetization can spontaneously emerge at the surface of an antiferromagnet with suitable symmetry, including the (110) surface of an AFM with a rutile structure, such as FeF\(_2\), which is isostructural to RuO$_2$. Accordingly, surface magnetization is intrinsically linked to the magnetic structure of the bulk. However, as shown above, surface magnetization also arises from the local breaking of bulk symmetry at the surface of a non-magnetic material. Further, we want to emphasize that the surface magnetism seen in our simulation is robust with DFT functionals as long as the functional gives rise to non-magnetic bulk state, i.e., those PBE+$U$ with the on-site $U$ values not exceeding 1.06 eV~\cite{smolyanyuk2024fragility}. Thus, the presence of surface magnetization alone does not constitute necessary or sufficient conditions to infer the bulk magnetic structure. Bulk measurements are required to identify the magnetic domain conclusively.

%\subsubsection{The origin of surface magnetization in RuO$_2$(110) surface}
\subsubsection{The cause of surface magnetization in RuO\texorpdfstring{$_2$}{2}(110)}

\begin{figure}
\centering
\includegraphics[width=3.4in]{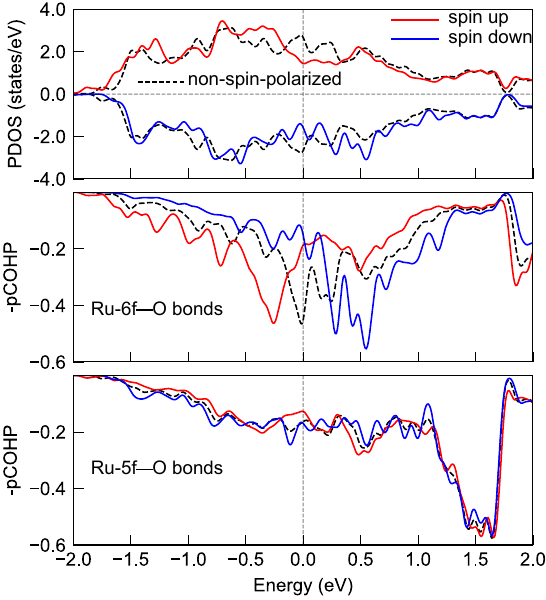}
\caption{Chemical bonding analysis for the emergence of surface magnetization. (a) Comparison of PDOS around Fermi level in the non-magnetic state, and spin-polarized state of the surface, (b) \(-\)pCOHP curves averaged for Ru--O bonds around the Ru-6f (Ru-center), and (c) around the Ru-5f (Ru-corner) atoms at the surface ($E_{\text{F}}$ set to 0).}
\label{fig3}
\end{figure}

In this section, we investigate the origin of surface magnetization on RuO$_2$(110), considering the non-magnetic bulk state. From bulk RuO$_2$ to the (110) surface, the symmetry of the already distorted octahedral around the Ru-corner and Ru-center atoms of the surface layers is further reduced due to surface termination. The Ru-corner (Ru-5f) loses one of its axially coordinated oxygen ligands, while the Ru-center (Ru-6f) maintains its coordination number. The two bridging oxygen ligands on the surface, which coordinate equatorially to the Ru-6f atom, become covalently \textit{unsaturated} with only two bonds remaining as compared to three bonds in the bulk, contributing to structural distortions that affect the hybridization between Ru $4d$ and O $2p$ states. These distortions are expected to lift the degeneracy of the Ru $4d$ states further. Without any surface relaxation, i.e., no structural relaxation performed for slabs, the local symmetry of the Ru-5f is $C_{4v}$ (square pyramidal), while the Ru-6f retains its distorted octahedral symmetry as in the bulk. Nevertheless, this configuration is unstable, leading to structural relaxation akin to a Jahn-Teller distortion, which further lowers the local symmetry around the surface Ru atoms. During this relaxation, the Ru--O axial bonds in the Ru-5f-centered square pyramidal shorten, whereas the axial bonds in the Ru-6f-centered octahedra lengthen. Additionally, the equatorial Ru--O bonds centering around the Ru-6f become asymmetrically distorted; those involving the bridging oxygen ligands are significantly shortened, while the others elongate slightly. Despite these structural reconstructions, the total energy of the system remains high, with a considerable DOS at the Fermi level contributing to the occupation of anti-bonding states, as illustrated in Fig.~\ref{fig3}(a).

To achieve stability, spin polarization is necessary to redistribute electronic densities of the two spin sublattices, resulting in spontaneous surface magnetization as shown in the previous section. Crystal orbital Hamilton population (COHP) analysis offers valuable insight into the driving forces behind the emergence of magnetization from a chemical bonding perspective~\cite{ref39Landrum2000,ref40Dronskowski2004}. Fig.~\ref{fig3} presents the projected density of states (PDOS) on the atoms of the top surface, along with \(-\)pCOHP curves (averaged for the nearest-neighbor Ru--O contacts) for both structurally relaxed non-magnetic and magnetic states. In the non-magnetic state, despite the structural relaxation, RuO$_2$(110) displays a peak at the Fermi level in both the PDOS and \(-\)pCOHP (with negative values) curves, indicating an unstable configuration. This suggests that structural effect alone is not enough to stabilize the surface system. Consequently, the surface exhibited pronounced electronic reorganization upon spin-polarization calculation. The surface atoms become magnetized and the electronic structure shows substantial spin-polarization [Fig.~\ref{fig2}(e) and ~\ref{fig3}(a)]. Remarkably, this spin-polarized reorganization shifted the DOS and \(-\)pCOHP peaks away from the Fermi level, leading to a drastic decrease in the density of antibonding states at the Fermi level for both spin channels, and effectively stabilizing the system by about 43 meV. The electronic redistribution effect is stronger for Ru-6f compared to that of Ru-5f, thus giving rise to larger moments on the former as seen in the previous section. In addition, when performing further structural relaxation with the inclusion of spin-polarization, we observed no discernible structural difference to the non-spin-polarized case, gaining only 2 meV in total energy. Ultimately, it is the surface spin-polarization that stabilizes the surface by about 45 meV per unit cell, regardless of the slab thickness, as mentioned in the previous section.

The development of surface magnetization on RuO$_2$(110) can also be understood through the lens of the Stoner criterion, which states that magnetization emerges if the product of the density of states at $E_{\text{F}}$ and the exchange interaction parameter exceeds unity~\cite{ref47stoner1938collective}. In the non-magnetic state, even after the structural relaxation, the high DOS at the Fermi level suggests the surface instability. This unstable state drives the surface to electronic redistribution, and the system minimizes its energy through spin polarization. This reorganization substantially lowers the DOS for both spin channels around $E_{\text{F}}$ as described above, stabilizing the surface. The significant reduction in antibonding states and the redistribution of electronic density through spin polarization effectively meet the Stoner criterion, leading to spontaneous surface magnetization.

Electronic reconstruction that results in the emergence of induced magnetic moments and net magnetization in the surface layer can be further seen in the variation of charge states of individual atoms at the surface compared to their bulk counterparts, as detailed in Table~\ref{tables2}. Different population methods yield varying values for these charge states, but the trend of electronic reconstruction for the surface atoms remains consistent across these methods. For example, using the charge density-based Bader approach, we find that the charge state of the Ru-5f atom generally decreases, while that of the Ru-6f atom increases relative to the bulk values. This is primarily due to the absence of one oxygen atom coordination (the axial oxygen) of the Ru-5f, which leads to an increased electron density and a corresponding decrease in charge state at the Ru-5f site. In contrast, the Ru-6f retains full coordination with six oxygen ligands, but also specially interacts with two bridging oxygen ligands having higher hole densities due to one unsaturated coordination for each oxygen. This condition facilitates charge transfer from the Ru-6f to these covalently unsaturated ligands, thereby increasing its charge state.

%\subsubsection{The significance of surface magnetism in RuO$_2$(110) surface}
\subsubsection{The significance of surface magnetism in RuO\texorpdfstring{$_2$}{2}(110)}

\begin{figure*}
\centering
\includegraphics[width=5.0in]{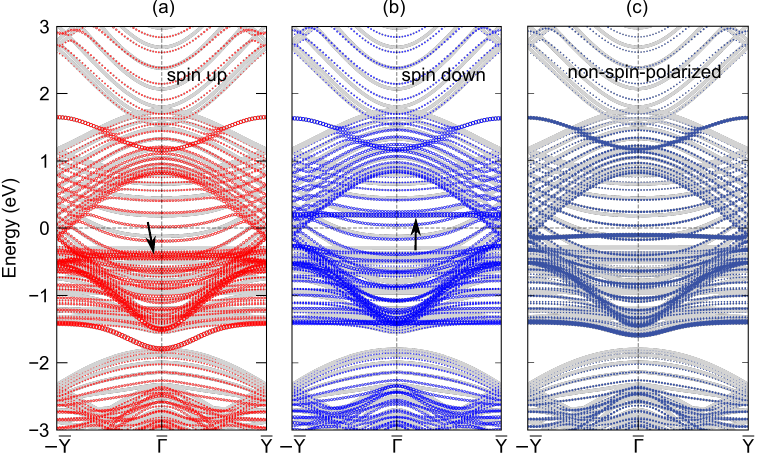}
\caption{Projected electronic band structure for the atoms of the outermost layer plotted along the corresponding projected bulk bands shown in the light-gray background ($E_{\text{F}}$ set to 0). Each layer contains 2 formula units of RuO$_2$. (a) Majority spin (spin up, red), (b) minority spin (spin down, blue), (c) non-spin-polarized band. Surface bands can be seen with high intensity of weighted data, those near the Fermi level pointed to by black arrows.}
\label{fig4}
\end{figure*}

Surface magnetism developed on the surface of RuO$_2$(110) suggests that we can probe this phenomenon directly, i.e., the spin-splitting effect associated with surface bands should be detectable by surface- and spin-sensitive spectroscopy techniques such as spin-resolved ARPES. Fig.~\ref{fig4} shows the slab band structure with weighted contribution from the top layer atoms (2 Ru and 4 O, cf. Fig.~\ref{fig2}(d)) along with projected bulk bands in the light-gray colored background. Interestingly, low energy electronic structures close to the Fermi level exhibit surface states with flatness and significant spin-splitting exchange energy of approximately 0.60 eV, as indicated by black arrows in Fig.~\ref{fig4}(a) and ~\ref{fig4}(b). The occupied flat bands, which lie approximately 0.40 eV below $E_{\text{F}}$, are derived from all magnetic atoms of the surface layer, i.e., Ru-5f, Ru-6f, and O-br, with the strongest contribution from Ru-6f followed by O-br and the lowest contribution from Ru-5f, which is consistent with previous observation~\cite{ref48Jovic2021} (cf. Fig.~\ref{figs3} in SM). Other non-magnetic atoms of the top layer such as O-3f and O-sub contribute negligibly to these states. The minority spin counterparts of these flat bands are unoccupied (located ~0.20 eV above $E_{\text{F}}$) and dominated by the Ru-6f character and a smaller contribution from O-br (cf. Fig.~\ref{figs3} in SM). This is consistent with the highest calculated magnetic moment value for the Ru-6f atom among others on the surface.

It should be noted that the presence of occupied flat bands near the Fermi level has been observed in previous experiments by using ARPES~\cite{ref8liu2024,ref22Lin2024,ref49Jovic2018,ref50Jovic2019}. However, the spin polarization of these bands and the presence of their spin exchange counterpart above $E_{\text{F}}$ have not been reported in the literature. This is likely due to the ARPES technique probing only occupied states. Therefore, unoccupied states-sensitive experiments such as inverse ARPES, scanning tunneling microscopy (STM), or X-ray absorption spectroscopy (XAS) might be required to realize the full spin-polarized spectrum of the flat bands in the proximity of the Fermi level due to the surface magnetism in RuO$_2$(110).

As mentioned above, the existence of flat bands lying closely beneath the Fermi level has been observed experimentally, but the origin of these states is still a debate in the literature. Jovic {\em et al.}~\cite{ref50Jovic2019} showed that the flat bands have a surface origin. They were considered to be topologically trivial states that connect the projection of nodal lines along the XR directions in the bulk BZ~\cite{ref50Jovic2019}. Very recently, based on spin-resolved ARPES data, Liu {\em et al.}~\cite{ref8liu2024} assigned the flat bands near the Fermi level along the $\Gamma$--$\mathrm{M}$ momentum direction as surface states. However, a recent experiment has also shown the existence of flat bands along that $\Gamma$--$\mathrm{M}$ direction, but the authors argued that it likely has a bulk origin~\cite{ref22Lin2024}. It is noteworthy that our projected bulk spectrum exhibits a few flat bands residing below the Fermi level and overlapping in energy with the flat bands of interest, i.e., those with significant contributions from the outermost layers as described above. This overlap could be the source of confusion in the literature.

To reconcile the origin of these flat bands, we calculated the layer-resolved (projected) band structure for each layer of the slab, and the results are shown in Fig.~\ref{figs4} of SM. The layer-projected band structures show a rapid decrease in the contribution of each layer to the flat bands of interest when going from the outermost to the innermost (or the middle, the bulk-like) layer. The middle layer of the slab displays no fingerprint of the pairs of bands being discussed (Fig.~\ref{figs4}(f)), suggesting that these bands are derived from the top and bottom layers (the surface layers). In addition, the charge density associated with the flat bands at the \(\overline{\Gamma}\) point plotted in Fig.~\ref{figs5} clearly shows the outermost-layer origin of the bands. The fact that those surface bands are spin-split is a clear evidence for the emergence of surface magnetization mostly due to the atoms at the surface. This is consistent with the observation that the developed local moments on atoms decay quickly when going from the surface to the bulk region.

Another pair of spin-splitting bands can be observed around \(-\)1.5 eV where the minority spin channel is completely flat while the majority counterpart is slightly dispersive. This contrasts with the corresponding surface-dominated bands of the non-magnetic surface where these bands are also slightly dispersive in the proximity of the \(\overline{\Gamma}\) point but without spin polarization (Fig.~\ref{fig4}(c)). Similar to the flat band surface states close to the Fermi level, these spin-polarized states split in energy, i.e., one moving up and the other moving down compared to non-magnetic states upon spin-polarization calculation. In addition, there is another set of surface states sitting at around 1.5 eV above the Fermi level. These states are slightly dispersive near the \(\overline{\Gamma}\) point and do not show a significant exchange splitting, likely due to the dominating contribution of the small moment Ru-5f atoms to the band composition.

Considering that the bulk RuO$_2$ is non-magnetic and our calculations exclude relativistic effects, the emergence of significant spin exchange splitting associated with surface bands near the Fermi level underscores the unique feature of the spontaneous surface magnetization of RuO$_2$(110). This characteristic is likely a key factor in determining the magnetic properties of surface RuO$_2$. Additionally, the presence of large spin-splitting surface bands, particularly those close to the Fermi level, suggests that the spin polarization of transport at the surface would differ significantly from that in the bulk, which could profoundly influence spin-dependent phenomena in RuO$_2$. Therefore, due to the induced exchange interactions, proximity effect, and potential spin current manipulation at the surface, we conjecture that surface magnetism could play a vital role in the material’s spin Hall magnetoresistance when coupled with heavy metals such as Pt or Ta~\cite{kobayashi2024detection}, or serve a relevant source for interfacial exchange coupling when paired with a magnetic material.

\begin{figure}
\centering
\includegraphics[width=3.4in]{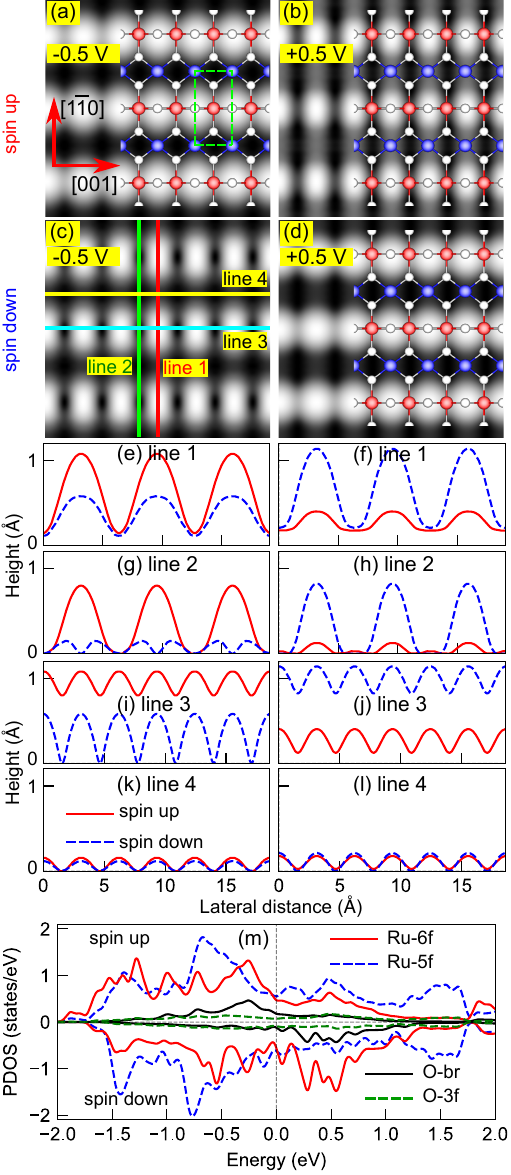}
\caption{Simulated STM images of (110) surface. (a)--(d) Images at bias voltages of $-0.5$ V (left column) and $+0.5$ V (right column); the unit cell of the surface highlighted by a dashed green rectangle; Ru-6f, Ru-5f and oxygen atoms represented by red, blue, and bright gray balls, respectively. (e)--(l) Bias- and spin-dependent line profiles indicated as lines 1, 2, 3, and 4 in (c), left and right column figures corresponding to $-0.5$ V and $+0.5$ V bias voltages, respectively. (m) Site-projected DOS of surface atoms explaining the difference in spin-dependent images. Atomic labels denoted as in Fig.~\ref{fig2}(b).}
\label{fig5}
\end{figure}

Surface magnetization can also exhibit different characteristics when probed by surface- and spin-sensitive microscopy techniques such as spin-polarized scanning tunneling microscopy (SP-STM)~\cite{ref51Heinze2000}. Since SP-STM is applicable to electrically conductive materials, such as magnetic metals or doped semiconductors, RuO$_2$’s metallic nature makes it a suitable candidate for the technique. General observation from STM images shown in Fig.~\ref{fig5} is the high intensity at bright spots corresponding to the O-br--derived electronic states due to their positions lying closest to the tip in STM measurement. This has been theoretically simulated and experimentally observed~\cite{ref52Over2004,ref53Feng2021}. Notably, although Ru-5f exhibits a high DOS for both spin channels below the Fermi level, the signal is predominantly dominated by bridging oxygen atoms rather than Ru-5f [(cf. Fig.~\ref{fig5}(a),(c) and Fig.~\ref{figs6}(a),(b),(e),(f)]. This occurs because the Ru-5f atoms are located one atomic layer further from the STM tip compared to the bridging oxygen atoms, resulting in an exponential decrease in tunneling current with distance. Furthermore, the occupied states of Ru-5f originate from in-plane Ru $4d$ orbitals (Fig.~\ref{figs6}(i)), which do not extend significantly out of plane. This interpretation is further supported by spin density visualizations shown in Fig.~\ref{figs6}(j),(l), revealing that, under negative bias voltage, the spin density of Ru-6f and O-br covers a greater spatial extent than that of the Ru-5f.

More importantly, our simulated STM images shown in Fig.~\ref{fig5} reveal clear differences in the spin-dependent images. For instance, under negative bias voltage (i.e., probing \textit{occupied} states), we can see high protrusions at Ru-6f and O-br positions forming horizontal lines in the spin-up images. This is due to the occupied states being dominated by the majority spin (spin-up) channel mostly derived from Ru-6f and O-br as identified by the atomic (site-) projected density of states for surface atoms (Fig.~\ref{fig5}(m)) and the charge distribution from partial charge density calculations (Fig.~\ref{figs6}(j)). In contrast, when probing the \textit{unoccupied} states under positive bias voltages, higher protrusions forming horizontal lines in the obtained images were derived from the spin-down channel.

The spin-dependent feature of the STM images can be clearly corroborated further by the height--profiles of the signals scanned over four selected lines indicated in Fig.~\ref{fig5}(c). The profile lines displayed in the left-column figures [Fig.~\ref{fig5}(e),(g),(i),(k)] corresponding to the negative bias voltage images exhibit periodic patterns consistent with the dominant contribution of the spin-up density from Ru-6f and O-br positions. In contrast, the right-column figures [Fig.~\ref{fig5}(f),(h),(j),(l)], representing positive bias voltage scans, show the significant effect of the spin-down density from the same atomic positions. This aligns well with the expected contribution of the majority and minority spin density of states in a spin-polarized system.

An intriguing feature emerges when transitioning from negative to positive bias: higher positive biases enhance the visibility of the Ru-5f topography. At +0.5 V bias, the position of Ru-5f becomes discernible in the spin-up images; and at +1.0 V bias, bright vertical columns along [\(1\overline{1}0\)] appearing periodically along the [001] crystallographic direction are obvious (Fig.~\ref{figs6}(c)). This effect arises from the dominant contribution of the Ru-5f spin-up channel over the spin-down channel in its \textit{unoccupied} states as demonstrated in Fig.~\ref{fig5}(m) and Fig.~\ref{figs6}(i). As the bias voltage increases, the position of Ru-5f atoms become more clearly visible due to the higher contribution from the out-of-plane components of its $4d$ orbitals as shown in Fig.~\ref{figs6}(i), exhibiting a phenomenon known as contrast reversal. This behavior has previously been observed in studies of rutile transition-metal oxide (110) surfaces and helps distinguish stoichiometric surfaces from other structures under varying experimental conditions, further validating our STM image simulations~\cite{ref53Feng2021}.

%%%%%%%%%%%%%%%%%%%%%%%%%%%%%%%%%%%%%%%%%%%%%%%%%%%%%%%%%%%%%%%%%%%%%%%%%%%%%%%%%%%%%%%%%%%%%%%%%%%%%%%%%%%%%%%%%%%%%%%%%%%%%%
\section{Conclusions}\label{sec:Conclusion}

In summary, we have shown by first-principles calculations that surface magnetization emerges from a metallic non-magnetic bulk material. The magnetization is significantly large at the outermost layers of the slab structure and decays quickly as it gets deeper into the bulk region. Symmetry lowering as a result of surface termination induces not only structural relaxation but also significant electronic reorganization at the surface layer, resulting in sizable moments on the atoms forming the surface. This spontaneous surface magnetism proved to exhibit unique properties, including spin-split surface bands near the Fermi level and distinct spin-resolved spectroscopic features. These findings are expected to provide valuable insights into the interpretation of experimental observations in spectroscopy and spin-dependent transport phenomena, highlighting the pivotal role of surface magnetism in such systems.

%%%%%%%%%%%%%%%%%%%%%%%%%%%%%%%%%%%%%%%%%%%%%%%%%%%%%%%%%%%%%%%%%%%%%%%%%%%%%%%%%%%%%%%%%%%%%%%%%%%%%%%%%%%%%%%%%%%%%%%%%%%%%%
\section{Acknowledgement}
We thank D. Wickramaratne, J. Xiao, S. Bhatt, and D. Plouff for fruitful discussions. This work was supported by the NSF through the UD-CHARM University of Delaware Materials Research Science and Engineering Center (MRSEC) Grant No.~DMR-2011824. A.J. acknowledges support from the U.S. Department of Energy (Contract No.~DE-SC0014388). We also acknowledge the use of Stampede3 at TACC through allocation PHY240154 from the Advanced Cyberinfrastructure Coordination Ecosystem: Services \& Support (ACCESS) program, which is supported by National Science Foundation Grants No.~2138259, 2138286, 2138307, 2137603, and 2138296, and the DARWIN computing system at the University of Delaware, which is supported by the NSF Grant No.~1919839.
% \blindtext \cite{article-minimal}
\bibliographystyle{apsrev4-1} % Tell bibtex which bibliography style to use
\bibliography{main.bbl} % Tell bibtex which .bib file to use (this one is some example file in TexLive's file tree)

\begin{thebibliography}{61}
\bibitem [{NIS()}]{NISTdisclaimer}%
  \BibitemOpen
  \href@noop {} {}\bibinfo {note} {Certain commercial equipment, instruments, software or materials are identified in this paper to specify the procedures or equipment used. Such identification is not intended to imply recommendation or endorsement by the National Institute of Standards and Technology, nor is it intended to imply that the materials, equipment or software identified are necessarily the best available for the purpose.}\BibitemShut {Stop}%
\end{thebibliography}

%%%%%%%%%%%%%%%%%%%%%%%%%%%%%%%%%%%%%%%%%%%%%%%%%%%%%%%%%%%%%%
\clearpage % To start a new page for Supplemental Material
\onecolumngrid
\section{Supplemental Material}

% Change figure labeling to "S" prefix for Supplemental Material
\renewcommand{\thefigure}{S\arabic{figure}}
\setcounter{figure}{0} % Reset figure counter
\renewcommand{\thetable}{S\arabic{table}}
\setcounter{table}{0} % Reset table counter for "S" numbering
%% Table S1
\begin{table*} [h]
    \centering % Center the first table
    \setlength{\tabcolsep}{12pt}
    \caption{\(-\)IpCOHP values for Ru--O bonds in non-magnetic bulk RuO$_2$}
    \begin{tabular}{l c c c}
        \hline\hline
        Bond & Bond distance (\AA) & \(-\)IpCOHP value at $E_{\text{F}}$ \\ 
        \hline 
        Ru1--O3 & 1.941 & 4.034 \\
        Ru1--O4 & 1.941 & 4.035 \\
        Ru1--O5 & 1.984 & 3.486 \\
        Ru1--O6 & 1.984 & 3.487 \\ [0.5ex]
        \hline\hline
    \end{tabular}
    \label{tables1}
\end{table*}

%% Table S2
\begin{table*} [h]
    \centering % Center the first table
    \setlength{\tabcolsep}{12pt}
    \caption{Variation of atomic charge states of surface Ru and O atoms due to local symmetry breaking at the surface under the surface termination}
    \begin{tabular}{l c c c c c}
        \hline\hline
        Population & Ru-5f & Ru-6f & O-br & O-3f & O-sub \\  
        \hline 
        Bader    & 1.599 & 1.806 & -0.746 & -0.888 & -0.898 \\
        Mulliken & 1.390 & 1.520 & -0.610 & -0.750 & -0.760 \\
        Löwdin   & 1.190 & 1.320 & -0.540 & -0.640 & -0.670 \\ [0.5ex]
        \hline\hline
    \end{tabular}
    \label{tables2}
\end{table*}

%% FIG_S1
\begin{figure*}
\centering
\includegraphics[width=3.4in]{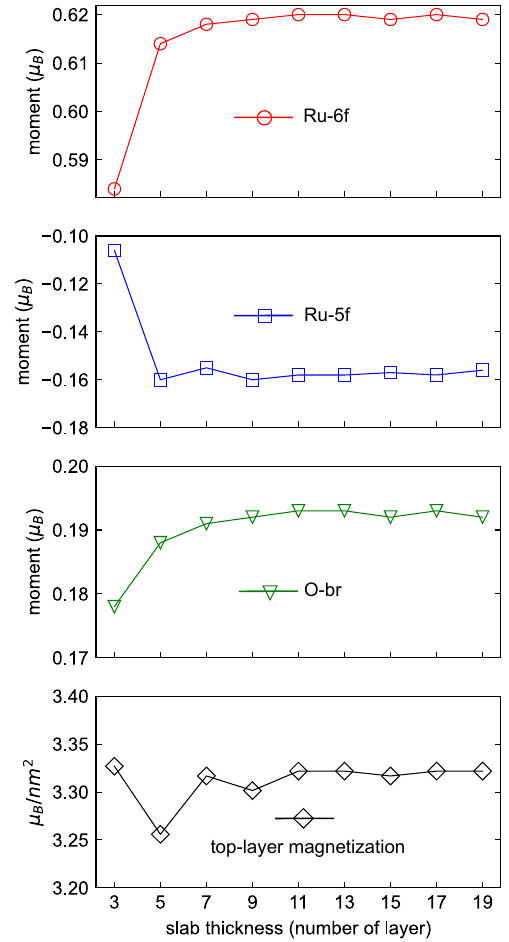}
\caption{Magnetic moments on magnetic atoms of the outermost-layer, i.e., the top or the bottom layer (top three panels), and its magnetization as a function of the slab thickness (bottom panel).}
\label{figs1}
\end{figure*}

%% FIG_S2
\begin{figure*}
\centering
\includegraphics[width=5.0in]{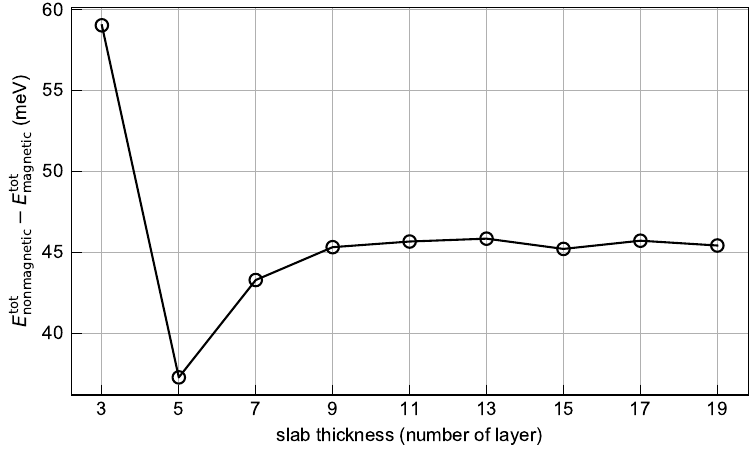}
\caption{Relative energy of spin-polarized state compared to non-magnetic state as a function of slab thickness \((E_{\text{nonmagnetic}}^{\text{tot}} - E_{\text{magnetic}}^{\text{tot}}, \, \text{meV})\), showing a more stable magnetic phase by approximately \(45 \, \text{meV}\).}
\label{figs2}
\end{figure*}

%% FIG_S3
\begin{figure*}
\centering
\includegraphics[width=7.0in]{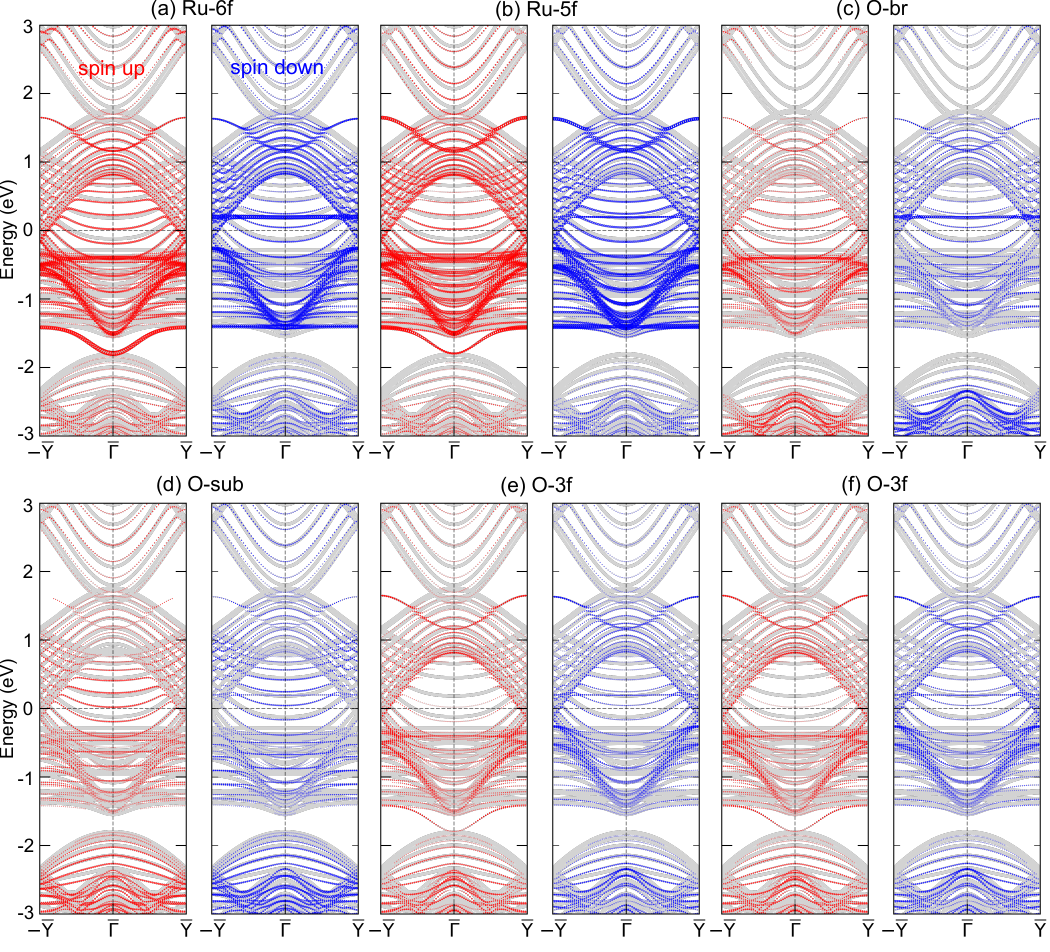}
\caption{Atom-resolved (projected) band structures for surface atoms showing the significant contribution of the bridging oxygen ligands and the two Ru atoms on the surface ($E_{\text{F}}$ set to 0). Atomic labels as denoted in Fig.~\ref{fig2}(b). Note that there are two equivalent O-3f atoms in the top layer, and their contributions to the band structures of the slab are shown separately in (e) and (f).}
\label{figs3}
\end{figure*}

%% FIG_S4
\begin{figure*}
\centering
\includegraphics[width=7.0in]{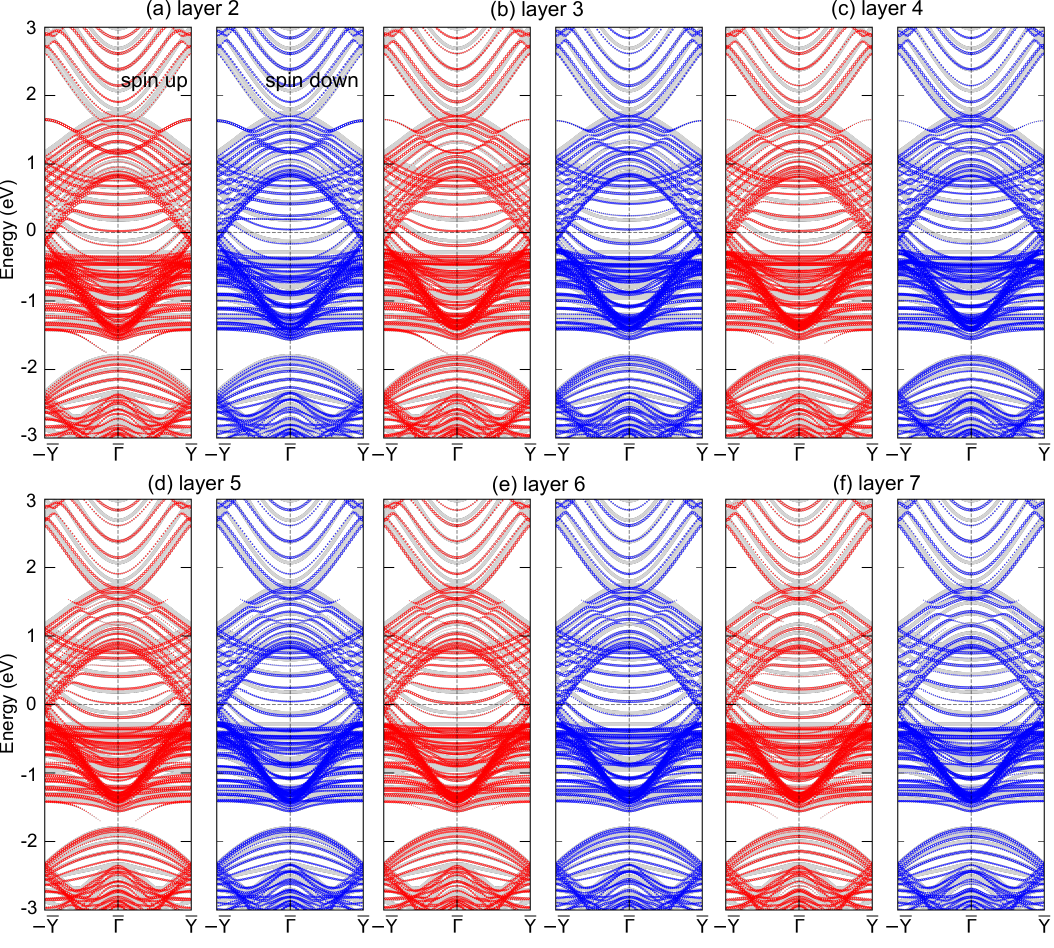}
\caption{Layer-resolved (projected) band structures for the 2nd to the 7th layer highlighting the decaying contribution to surface states when going deeper into the bulk-like region (layer 7th), $E_{\text{F}}$ set to 0. The labeling of layers can be found in Fig.~\ref{fig2}(d).}
\label{figs4}
\end{figure*}

%% FIG_S5
\begin{figure*}
\centering
\includegraphics[width=7.0in]{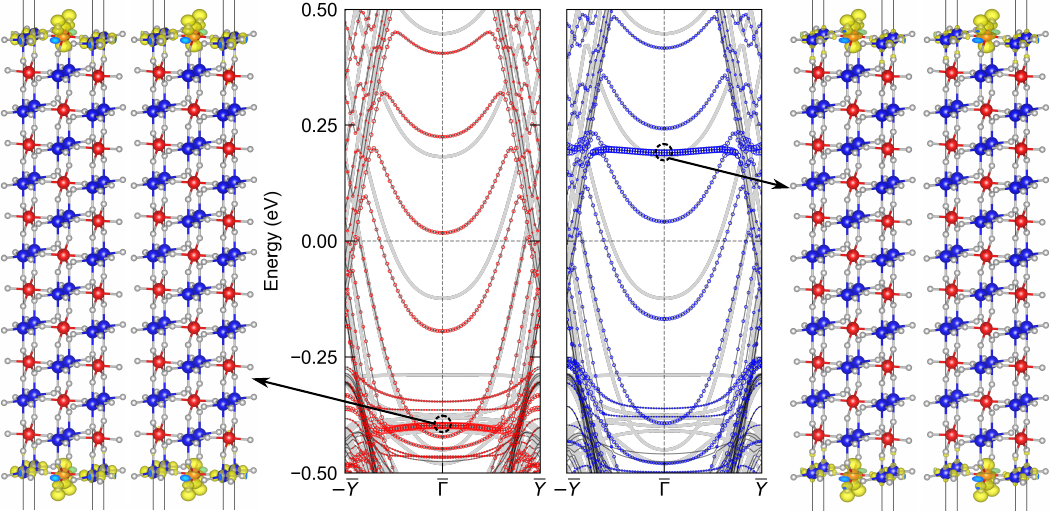}
\caption{Momentum and band-resolved spin densities of flat-band near the Fermi level which are derived from surface layers, the isosurfaces are plotted using 5\% of the maximum isosurface value. Spin densities shown on the left hand side are associated with the doubly occupied degenerate states, which can be seen as linear combinations of the orbitals of top and bottom surfaces. Similarly, the two minor spin degenerate states on the right hand side have contribution from outer layers only.}
\label{figs5}
\end{figure*}

%% FIG_S6
\begin{figure*}
\centering
\includegraphics[width=7.0in]{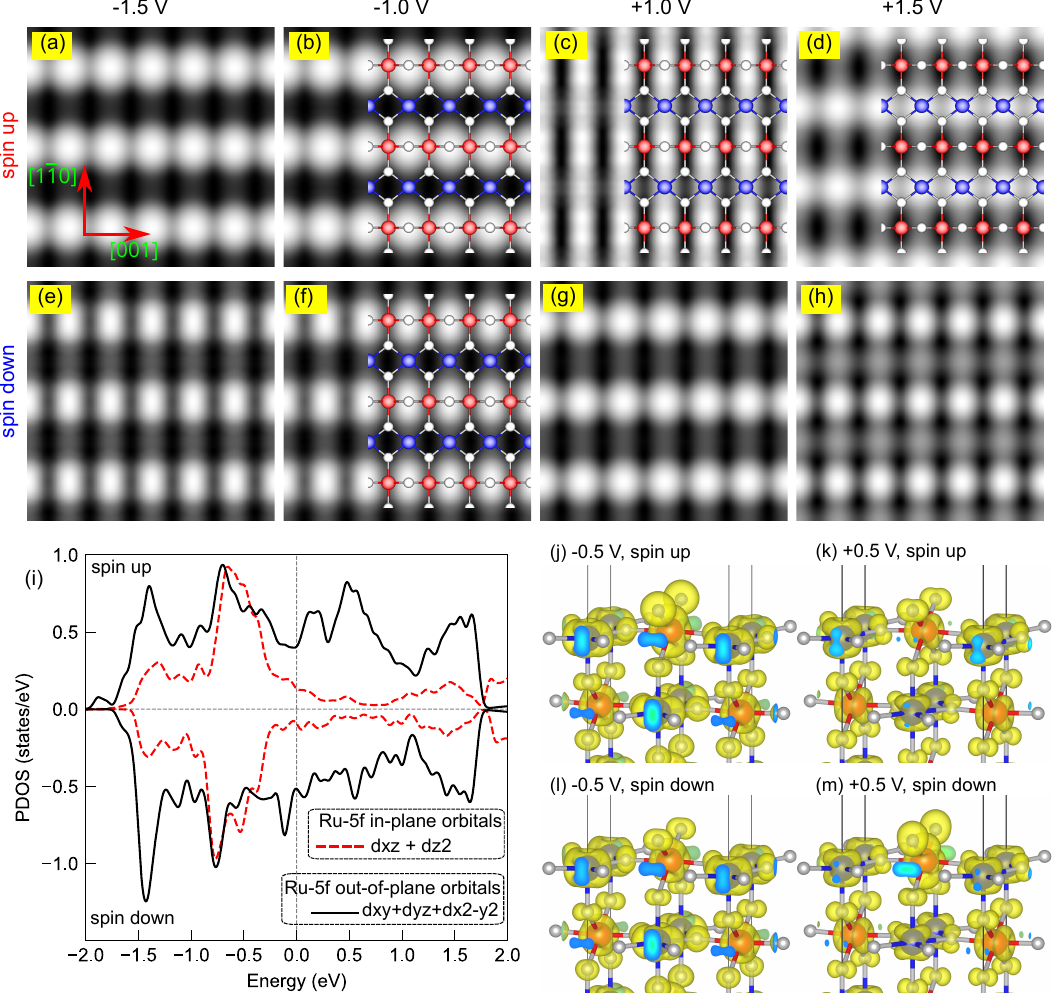}
\caption{Simulated STM images of (110) surface with various bias voltages ranging from \(-\)1.5 V to \(+\)1.5 V for spin up channel ((a) -- (d)) and spin down channel ((e) -- (h)), (i) the orbital-projected DOS of the Ru-5f atom, $E_{\text{F}}$ set to 0, and (j) -- (m) the spin-resolved charge densities seen in each corresponding STM simulation.}
\label{figs6}
\end{figure*}

\end{document}